# Ionic Kratzer bond theory and vibrational levels for achiral bond H$_2$

G. Van Hooydonk, Ghent University, Faculty of Sciences, Krijgslaan 281, B-9000 Belgium

Abstract. *A dihydrogen Hamiltonian reduces to the Sommerfeld-Kratzer-potential, adapted for field quantization according to old-quantum theory. Constants $\omega_e$, $k_e$ and $r_e$ needed for the H$_2$ vibrational system derive solely from hydrogen mass $m_H$. For H$_2$, a first principles ionic Kratzer oscillator returns the covalent bond energy within 0,08% and all levels within 0,02 %, 30 times better than Dunham's oscillator and as accurate as early ab initio QM.*

## I. Introduction

Physicists focused on the *simple line spectrum of atom H* with fine and hyperfine structure, less on the *complex band spectrum of molecule H$_2$* [1]. Since Bohr's simple, fairly accurate atom theory made H prototypical for atomic spectroscopy, a simple bond theory should make H$_2$ prototypical for molecular spectroscopy [2]. However, only complex QM theory accounts accurately for H$_2$ levels and its potential energy curve (PEC) [3,4]. With many parameters and hundreds of terms in the H$_2$ wave function [3], QM is far from transparent, which explains the success of DFT and illustrates why Bohr-type bond theories are still of interest [5].

Due to its complexity, QM fails on a simple analytical function for PECs and on a *low parameter universal function* (UF) [2,6,7], needed to unify shape-invariant, asymmetric PECs [2]. This failure justifies many attempts to find a UF [2,6], a problem usually assessed [2,6,7] with Dunham theory [8]. If H$_2$ were the best starting point to get at universal behavior, its levels must be understood with a simple low order potential like the Dunham or Kratzer oscillator [2,6].

Since anharmonicity flaws the harmonic oscillator (HO), so important for modern physics [9], we start with the HO, which we confront with the H$_2$ spectrum in Section II. Dunham and Kratzer oscillators are compared in Section III. In Section IV, old quantum theory leads to a quantized ionic Kratzer bond theory [2], whereby all H$_2$ parameters, $r_0$, $\omega_e$ and $k_e$, derive solely from mass $m_H$. Section V, on the accuracy of Dunham and Kratzer oscillators, proves that Kratzer theory is as accurate as *earlier ab initio QM methods* [10]. Discussions and conclusion are in Sections VI and VII.

## II. Quantum HO and anharmonicity in bond H$_2$

H$_2$ rotator-vibrator levels $E_{v,J}$ vary with vibrational and rotational quantum numbers v and J. For vibrational levels v (J=0), Schrödinger's quantum HO [11] gives equally spaced levels, according to

$$E_{v+\frac{1}{2}} = \omega_e(v+\tfrac{1}{2}) \text{ cm}^{-1} \text{ or } E_{v+\frac{1}{2}}/\omega_e = v+\tfrac{1}{2} \qquad (1a)$$

where $\omega_e$ is the fundamental vibrational frequency. Nevertheless, (1a) disagrees with the observed H$_2$ anharmonicity. A series expansion in half integer v

$$E_{v+\frac{1}{2}} = \omega_e(v+\tfrac{1}{2}) - \omega_e x_e(v+\tfrac{1}{2})^2 + \omega_e y_e(v+\tfrac{1}{2})^3 - \ldots \text{ cm}^{-1} \qquad (1b)$$

gives better agreement but this is equivalent with an expansion in integer v

$$E_v = A+Bv+Cv^2+Dv^3 - \ldots \text{ cm}^{-1} \qquad (1c)$$

Coefficients A, B, C… derive from those in (1b), e.g. $A=½\omega_e(1-x_e+y_e-…)$ cm$^{-1}$…. Fig. 1 gives the $E_v(v)$ plot for all 14 observed H$_2$-levels in Table 1 [12]. Empirical 1$^{st}$, 2$^{nd}$, 4$^{th}$ and 6$^{th}$ order fits give errors of respectively 1839,93; 111,84[1]; 7,15 and 0,24 cm$^{-1}$. Even a 6$^{th}$ order fit is not of *spectroscopic accuracy*. Ranking by accuracy places earlier ab initio QM [10], with errors of 3,2 cm$^{-1}$, in between 4$^{th}$ and 6$^{th}$ order fits. Errors of 1840 cm$^{-1}$ reveal that Schrödinger's famous HO formula (1a) fails for the simplest and stable vibrator in nature, H$_2$.

Using reduced mass, equilibrium separation $r_0$ and kinetic energy $T=½\mu v_0^2=½\mu\omega_e^2r_0^2$, vibrator energy $E_0=T_0+V_0$ depends on $V_0$, equal either to $-½k_er_0^2$ (Hooke) or to $-½e^2/r_0$ (Coulomb), using the virial. For $E_0=0$, fundamental frequency $\omega_e$ is available with 2 concurrent equations

$$\text{Hooke:} \quad E_0=½\mu_H\omega_e^2r_0^2-½k_er_0^2=0 \tag{1d}$$

$$\text{Coulomb:} \quad E_0=½\mu_H\omega_e^2r_0^2-½e^2/r_0=0 \tag{1e}$$

Hooke's law (1d) provides with the standard, classical HO relation

$$\text{Hooke :} \quad \mu_H\omega_e^2=k_e \text{ or } \omega_e=\sqrt{(k_e/\mu_H)} \tag{1e}$$

since $\mu\omega_e^2$ in $T_0$ is replaced by $k_e$ in $V_0$. Hooke's law does not really lead to a solution, since the problem is only shifted from $\omega_e$ to $k_e$, if mass $m_H$ (or $\mu_H$) is available.

Coulomb variant (1e) seems superior to (1d) as it gives an explicit solution for $k_e$, leading to

$$\text{Coulomb:} \quad k_e=e^2/r_0^3;\ \mu_H\omega_e^2=e^2/r_0^3 \text{ or } \omega_e=\sqrt{[e^2/(\mu_Hr_0^3)]} \tag{1g}$$

With $r_0=0,74$ Å for H$_2$, (1g) returns $\omega'_e=4380$ cm$^{-1}$, close to observed 4400 cm$^{-1}$ [12], a remarkable result, which led us to reconsider oscillator theory. If $r_0$ were available with classical physics, (1g) makes $\omega_e$ available with classical physics too. A classical value for $r_0$ would not only be a remarkable result, it would also restore the reliability and usefulness of classical physics, as we show below.

## III. Revisiting the HO: Dunham and Kratzer potentials

Sinusoidal solutions for HO (1a) derive from Hooke force $F=-k_er$ and Newton's 2$^{nd}$ law $F=ma$ [11]. With $V(r)=½k_er^2$, a Hooke-Dunham HO potential

$$V_{HO}=½k_e(r-r_0)^2=½k_er_0^2(r/r_0-1)^2=a_0d_D^2 \tag{2a}$$

*is so firmly entrenched that alternatives are rarely employed, even when it is known to be wrong* [2,15]: *it is only accurate for r close to $r_0$, it is symmetric instead of asymmetric in function of r and it can never converge*: it gives an infinity when $r \to \infty$. Dimensionless Hooke-Dunham variable

$$d_D = (r/r_0-1) \tag{2b}$$

transforms (2a) in $V_{HO}=a_0d_D^2$, where $V(r_0)=a_0=½k_er_0^2$. $V_{HO}/a_0=d_D^2$ has 2 solutions $\pm d_D$ for the r-dependence in non-convergent, symmetric PECs. Even Dunham's more flexible series expansion

$$V_{HO}=a_0d_D^2(1+a_1d_D+a_2d_D^2+…) \tag{2c}$$

identical with $V(r)=c_1(r-r_0)+c_2(r-r_0)^2+c_3(r-r_0)^3+c_4(r-r_0)^4+…$, still faces convergence problems [7].

---

[1] Morse's 2$^{nd}$ order $E_v=-161,113+4397,264v-128,187v^2$ cm$^{-1}$ [13] gives large errors of 112 cm$^{-1}$ (see [14]).

Alternative dimensionless Sommerfeld-Kratzer[2] variable [2]

$$d_{SK} = (1-r_0/r) = (1-1/d_D) \qquad (2d)$$

secures PECs are asymmetric and convergent *without expansions* [2,15]. Its oscillator [2,7,15]

$$V_{SK} = \tfrac{1}{2}k_e r_0^2 (1-r_0/r)^2 = a_0(1-r_0/r)^2 \qquad (2e)$$

reduced to $V_{SK}/a_0$, gives 2 solutions for asymmetric, convergent PECs, i.e. $\pm d_{SK}$. In this paper, we reveal which of the two potentials is the better choice to understand the $H_2$ spectrum.

## IV. First principles Bohr-type ionic Kratzer bond theory, based on old quantum theory

*IV.1 Classical total energy of the $H_2$ bond*

The total energy E (or Hamiltonian **H**) for 4-particle system $H_2$ in QM (with pairs of charge-conjugated leptons a,b and nucleons A,B) consists of 4 kinetic and 6 potential energy terms

$$\mathbf{H} = \tfrac{1}{2}p_a^2/m_a + \tfrac{1}{2}p_b^2/m_b + \tfrac{1}{2}p_A^2/m_A + \tfrac{1}{2}p_B^2/m_B - e^2/r_{aA} - e^2/r_{bB} - e^2/r_{bA} - e^2/r_{aB} + e^2/r_{ab} + e^2/r_{AB}$$

$$E = \tfrac{1}{2}m_a v_a^2 + \tfrac{1}{2}m_b v_b^2 + \tfrac{1}{2}m_A v_A^2 + \tfrac{1}{2}m_B v_B^2 - e^2/r_{aA} - e^2/r_{bB} - e^2/r_{bA} - e^2/r_{aB} + e^2/r_{ab} + e^2/r_{AB} \quad (3a)$$

Using n, v and J for electronic $E_{elec}$, vibrational $E_{vib}$ and rotational energies $E_{rot}$, E is equal to

$$E = E_{elec} + E_{vib} + E_{rot} = E_n + E_v + E_J$$

Since intra-atomic $E_{elec} = E_n$ is a constant $E_0$ for the $H_2$ vibrator and since rotational states $E_{rot} = E_J$ are not yet considered, (3a) is simplified by subtracting $E_0$ or

$$\Delta E = E_{vib} + E_{rot} = E - E_0 \approx +\tfrac{1}{2}m_A v_A^2 + \tfrac{1}{2}m_B v_B^2 - e^2/r_{bA} - e^2/r_{aB} + e^2/r_{ab} + e^2/r_{AB}$$

where $E_n = E_0 = \tfrac{1}{2}m_a v_a^2 + \tfrac{1}{2}m_b v_b^2 - e^2/r_{aA} - e^2/r_{bB}$.

For symmetric $H_2$, $r = r_{bA} = r_{Ba}$ and $mv^2 = m_A v_A^2 = m_B v_B^2$, with $m = 1836,15 m_e$ close to $m_H = 1837,15 m_e$, are appropriate and lead to

$$\Delta E = E - E_0 \approx 2(\tfrac{1}{2}mv^2) - 2e^2/r + (e^2/r_{AB})(1 + r_{AB}/r_{ab}) = mv^2 + (e^2/r_{AB})[1 + r_{AB}/r_{ab} - 2r_{AB}/r)]$$

$$\equiv mv^2 + (e^2/r_{ab})[(1 + r_{ab}/r_{AB} - 2r_{ab}/r)] \equiv mv^2 + (e^2/r)[(r/r_{AB} + r/r_{ab} - 2)]$$

wherein all composite Coulomb terms are *equivalent*. Magnitude and sign depend on the spatial configuration of the 4 particles in (3a) or on the distribution of the 4 unit charges in neutral $H_2$ [22]. A simple solution with an ionic approximation $-e^2/r_{AB}$ needs a negative sign, which requires that $2r_{AB}/r > 1 + r_{AB}/r_{ab}$ or that the charge distribution be inverted [22]. Also, ionic structures involve particle transfers[3], may lead to a parameter for kinetic energies and to a different $E_0$-value. Another simple solution is obtained when $r_{AB}$ is large, since plausible approximation $r_{AB} \approx r_{ab}$ gives

$$\Delta E = E - E_0 \approx mv^2 + 2e^2(1/r_{AB} - 1/r)$$

Division by 2, using equilibrium separation $r_0$ and $r_{AB} = r_1 = a_1 r_0$ and $r = r_2 = a_2 \cdot r_0$ returns

$$\tfrac{1}{2}\Delta E \approx \tfrac{1}{2}mv^2 + (e^2/r_0)(r_0/r_{AB} - r_0/r) = \tfrac{1}{2}mv^2 + (e^2/r_0)(r_0/r_1 - 1/r_2) = \tfrac{1}{2}mv^2 + (e^2/r_0)(1/a_1 - 1/a_2)$$

---

[2] Already in 1916, Sommerfeld used (2d) for H [16,17]. Prior to Schrödinger, his pupil Kratzer used it for a general bond theory [18]; his colleague Kossel [19] for an ionic bond theory. Fues [20] solved the wave equation for (2e).
[3] Attraction $-e^2/r$ typifies ionic bonding and a particle transfer between the neutral atoms X. With the large energy gap involved (equal to IE-EA, if IE and EA are ionization energy and electron affinity of atom X), such particle transfers, and therefore ionic bonding, are improbable at long range.

an extremely simple formal approximation $E_{vib}$. The ± sign for the composite Coulomb term can be due to a geometry dependent operator *parity* **P**, with $\mathbf{P}^2=1$ but also to a geometry independent *charge* operator **C**, with $\mathbf{C}^2=1$. However, **C** is forbidden, since it would transform an atom in an anti-atom [22]. Under these conditions, $E_{vib}$ for $H_2$ is therefore of general type

½$\Delta E \approx$ ½$mv^2+\mathbf{P}(e^2/r_0)(1/a_1-1/a_2)= [$½$mv^2+\mathbf{C}(e^2/r_0)(1/a_1-1/a_2)]$

all valid transformations for the vibrational part $E_{vib}$ of total energy E (3a), even when $m=m_H$. Our further analysis uses scaling by inter-nucleon separation $r_{AB}$, since (a) nucleons carry the greater moment of inertia; (b) this choice can lead to *an ionic approximation* to bonding; (c) $r_{AB}$ is important for the Born-Oppenheimer approximation and (d) it is the conventional variable for PECs. Coefficient $A_r$, referring to $r_{AB}$, transforms the vibrational part of (3a) in

$E_{vib}=\Delta E(=\Delta \mathbf{H}'=p^2/m \pm A_r e^2/r)=+mv^2 \pm A_r e^2/r_{AB}$

$A_r=1+r_{AB}/r_{ab}-2r_{AB}/r$ (3b)

This is an extremely simple dihydrogen Hamiltonian with 2 terms: T, with $m \approx m_H$ much larger than electron mass $m_e$, and composite V of Coulomb-type, which is algebraic. Coefficient $A_r$ is a *numerical form factor*, determined by the geometry (the configuration) of $H_2$, pending its charge distribution. If $A_r$ were a constant, deriving from a particular $H_2$ geometry, $E_{vib}$ (3b) is a formally simple but not a central Coulomb problem, since this pertains to $E_0$ or $E_{elec}$. The problem with the two operators **P** and/or **C**, leading to the algebraic form with ± for $A_r$ in (3b), is understood with classical physics. In fact, only attractive field $-A_r e^2/r_{AB}$ is consistent with the formation of a stable bond (attractive branch of a PEC for $r_{AB}>r_0$). An ionic process is, however, is *suspicious* for 2 *neutral atoms* but only when $r_{AB}>>r_{crit}$, i.e. above a critical distance, where *ionic and non-ionic* PECs would cross. At closer range, particle transfers[3] can occur and some specific interactions[4] in (3a) may interfere. Also m in an ionic approximation with $-e^2/r_{AB}$, where $r_{AB}$ is equal to the separation between 2 H atoms, leads to masses m, respectively equal to $m_{H+}=m_P=1836,153m_e$ and $m_{H-}=1838,153m_e$, the average of which is $m_H=1837,153m_e$, while reduced mass for ionic pair is hardly different (see below). By the same classical argument, repulsive field $+A_r e^2/r_{AB}$ is even more *suspicious* at long range, as it can never lead to stability [22]. In this view, $+A_r e^2/r_{AB}$ must be confined to the repulsive branch of a PEC for $r_{AB}<r_0$. With this classical view on the bond formation process, it is evident that the two signs in (3b) can no longer be neglected and must be considered. If so, an ambiguity remains with operators **P** and **C**, one of which must be responsible for this switch. We now proceed with (3b). Since linear periodical harmonic motion along 1D field axis r bears on a formal connection with rotations[5] (specified in Section IV.2), angular velocity ω leads to velocity v, defined as

$v=\omega r$ (3c)

---

[4] So-called Coulomb problem $+e^2/r_{ab}$, i.e. electron-electron repulsion, leads to computational difficulties in QM [24].
[5] As in Kratzer theory, the rotational frequency ω for a bond follows $\hbar/\mu r_0^2$, since momentum $p=mv$ is quantized using Bohr's hypothesis $mvr=pr=n\hbar$ and $p^2=(n\hbar/r)^2$ (a thesis confirmed by Compton and de Broglie).

Since T>0, plugging (3c) in (3b) gives two possibilities

$$\Delta E' = +m_H \omega^2 r^2 \pm A_r e^2/r \qquad (3d)$$

Conceptually, (3d) is related to (1e) and formally consistent with $H_2$ result (1g) in Section II. Solution $-A_r e^2/r$ for (3d) gives a vibrator equation $\omega^2 \sim (1/m_H r^3)$, just like (1g). It also returns Kepler's 3$^{rd}$ law for rotations, e.g. $\omega^2 r_0^3 = e^2/(\frac{1}{2}m_H) = e^2/\mu_H = C$ (for planetary orbital motion, C is related to GM, where G is Newton's constant). This second ambiguity with (3d) for rotational or vibrational motion remains with Hamilton's $p = mv = m\omega r$. The more specific solution for *a stable Coulomb system*, i.e. when $-A_r e^2/r = -e^2/r$, can now be confronted with old quantum theory, giving

$$p = f\hbar/r$$

if a field factor f is allowed for. For a Coulomb system, Hamiltonian (3d) transforms in

$$\Delta \mathbf{H'} = +p^2/m_H - e^2/r = (f^2\hbar^2/m_H)/r^2 - e^2/r \qquad (3e)$$

i.e. the Kratzer *Coulomb* potential [18]. Before using it for a $H_2$ bond theory, we discuss some consequences of (3e) by rewriting it as momentum equation $p = \pm\sqrt{[m_H(\Delta E' + e^2/r)]}$. In this view, a state for which $\Delta E' = 0$ returns a consistent value for the momentum equal to

$$p_0 = mv_0 = m\omega_0 r_0 = \pm\sqrt{(m_H e^2/r_0)}$$

For $H_2$, this $p_0$ value is confirmed by experiment, see (1g) and Section II, which, in turn, validates (3e). However, the same procedure with $+e^2/r$ for (3e) gives momentum $p = \pm\sqrt{[m_H(\Delta E' - e^2/r)]}$. For the same state $\Delta E' = 0$, $p'_0$ is imaginary, since

$$p'_0 = \pm\sqrt{[m_H(-e^2/r_0)]} = \pm i\sqrt{(m_H e^2/r_0)}$$

An *imaginary solution* for momentum, adhered to by Schrödinger and typical for wave mechanics, is *suspicious* since, with classical physics, $+e^2/r$ can never give a stable system [22]. This analysis shows why wave mechanical bond theories can be (unnecessarily) complex [22] (see Section I).

Given the importance of momentum for the theory of chemical bond $H_2$ as described by its band spectrum, we expand on the link between $p = m_H v$ and $\hbar/r$ using Bohr theory. We skip details and give the typical equations. (i) The 1$^{st}$ derivative $d/dr$ of (3d) gives forces $2m_H\omega^2 r$ and $A_r e^2/r^2$, securing that at $r_0$, $\omega = \omega_e$, $2m_H\omega_e^2 r_0^3 = 2m_H v^2 r_0 = A_r e^2$ and $E'_0 = -m_H\omega_e^2 r_0^2 = -\frac{1}{2}A_r e^2/r_0 = -A_r e^2/(2r_0)$ are the same formal classical virial results, obtained by Bohr for a rotating electron in atom H. (ii) The 1$^{st}$ derivative $d/d\omega$ gives $2m_H\omega r^2 = 2m_H vr$. Following Bohr, this is equal to an equi-dimensional constant of action (say Planck's $\hbar$), with a field scale factor f, as above. (iii) Relations for v and r are obtained by dividing corresponding terms in (i) and (ii), conform Bohr theory for the rotating electron[6].

(i)     $2m_H\omega^2 r^3 = 2m_H v^2 r = A_r e^2$ and $v^2 = A_r e^2/(2m_H r)$

(ii)    $2m_H\omega r^2 = 2m_H vr = 2p_H r = f\hbar$ and $v = f\hbar/(2m_H r)$           (3h)

(iii)   $v = A_r e^2/(f\hbar)$ and $r = \frac{1}{2}f^2\hbar^2/(A_r m_H e^2)$

Using (ii) in (3h), gives $T = p_H^2/m_H = (f^2\hbar^2/m_H)/r^2$ and plugging this in (3d), the Hamiltonian for $H_2$ is

$$\Delta \mathbf{H'} = -A_r e^2/r + (f^2\hbar^2/m_H)/r^2 \qquad (3i)$$

---

[6] Here $v_e = e^2/n\hbar = \alpha c/n$, $r_B = \hbar^2/(\mu_e e^2)$ and $\alpha \approx 1/137,036$ is Sommerfeld's fine structure constant (see further below).

The 1st derivative of (3i) gives $½A_r e^2 r_0 = 2f^2\hbar^2/m_H$ and $\Delta\mathbf{H'} = -A_r e^2/r + ½A_r e^2/r^2$, i.e.

$$\Delta''\mathbf{H} = \Delta\mathbf{H'} + ½A_r e^2/r_0 = +(½A_r e^2/r_0)(1-r_0/r)^2 = V_{SK} \qquad (3j)$$

the Sommerfeld-Kratzer oscillator potential (2d) and (3j). Old quantum theory gives a transparent solution for the vibrational levels by means of a Kratzer oscillator, not a Hooke-Dunham oscillator. This proves why Kratzer's potential is indeed superior to Dunham's [2]. Since $V_{SK}$ is of closed self-contained analytical form, no other terms are required and a wave equation is not needed to get at vibrational energy $E_{vib}$ [22]. Differences between Coulomb models for atom H ($E_n$) and for bond $H_2$ ($E_n+E_v$) are the appearance of hydrogen mass $m_H$, field factor f and form factor $A_r$. However, solution (3j) can only be called classical, if $A_r$ and $r_0$ were available classically too (Section IV.2). The 2nd derivative $d^2/dr^2$ of (3d) for force constant equations $2m_H\omega_e^2$ and $2A_r e^2/r_0^3$ gives

$$\omega_e^2 = A_r e^2/(m_H r_0^3) \qquad (3k)$$

$$k_e = A_r e^2/r_0^3 \qquad (3l)$$

or, $k_e$ (3l) cannot be obtained *with the Hooke-Dunham oscillator for Schrödinger's HO* (1a), see Section II. [All solutions above have first principle's status. For diatomic bond $H_2$, reduced mass

$$\mu = m_H m_H/(m_H+m_H) = m_H/(1+m_H/m_H) = ½m_H \qquad (3m)$$

should be used indeed of mass $m_H$, as mentioned above, which is equivalent with using scale factor $s=½$ for dimer $H_2=H_AH_B$. In general, dimensionless recoil correction for bond AB ($m_A$, $m_B$)

$$s = 1/(1+m_A/m_B) \qquad (3n)$$

gives $s=½$ (3m) for dimers like $H_2$. All equations can be adapted accordingly. While for an electron in H, the ratio of mass and reduced mass is $\mu_e/m_e = 1+m_e/m_p \approx 1$, it is equal to ½ in $H_2$. A similar model dependent form factor can appear for I (moment of inertia for sphere, shell…, see below)]. With reference to Section II, it is tempting to associate unidentified $A_r e^2$ with $e^2$, which implies that the inter-atomic field is indeed of *ionic Coulomb-type* [23]. If valid,

$$k_e = A_r e^2/r_0^3 \equiv e^2/r_0^3 \qquad (3o)$$

(3o) also implies that, at $r_0$, *Coulomb attraction* $-e^2/r_0$, i.e. *ionic bond energy* $D_{ion}$, appears in *covalent* $H_2$ [2,23]. We recently found [23] that plugging observed $r_0=0,74$ Å [25] in (3o) returns observed $k_e=5,7.10^5$ dyne/cm and $\omega_e \approx 4400$ cm$^{-1}$ for $H_2$ [12,25], see Section II. This validates *unprecedented result* (3o) *a posteriori* as well as solution (3e) but the problem of assessing $r_0$ classically remains. Instead of borrowing this from experiment, we look for a classical solution for $r_0$ to safeguard the first principles, classical status of this theory (Section IV.2).

Solution (3o) for *ionic Coulomb bonding* at $r_0$ in *covalent bond* $H_2$, transforms (3j) further in [23]

$$\Delta\mathbf{H} = +(½e^2/r_0)(1-r_0/r)^2 = V_{SK} \qquad (3p)$$

Oscillator (3p) only derives from the *Hamiltonian substitution* p=mv and *old-quantum recipe* p=ℏ/r, originally due to *Bohr and Sommerfeld* and later conformed by *Compton* and *de Broglie*, whose works all appeared prior to *Schrödinger's*. This explains why wave equation and wave functions, generated by the Schrödinger interpretation of p, are not really needed [22]. Since PEC (3p) is *convergent and asymmetrical*, it is a convenient basis for a bond theory, although *ionic* potentials refer to old-fashioned *19th century ionic bonding theories* [2,19,22,23].

Whereas *not converging, symmetric Dunham oscillators* are typified with (2a) and variable $x=r/r_0$ (2f), *generic converging asymmetrical Coulomb Kratzer oscillators* obey (2d), (3h), (3p) and use inverse variable $1/x=r_0/r$. Coulomb oscillators are perfectly symmetrical and harmonic in variable $1/x$ instead of $x$. The result that $-A_r e^2$ is equal to $-e^2$ is surprising but that form factor $A_r$ is a *constant*, is even a greater surprise. We may have solved the $A_r$-algebra with old quantum theory and the Kratzer potential; we are still left with the problem of the $H_2$ geometry, imposed by $-A_r=-1$. Before doing so, we first try to find $r_0$ with classical physics, since a classical approach to $r_0$, i.e. to $\omega_e$ as in (1g), is important for the usefulness of classical physics for bonding (see Section II) and for the calculation of PEC (3p).

*IV.2 Vibrational frequency, equilibrium separation of a Coulomb vibrator and quantum hypothesis for $H_2$*

Since we need an independent calculus of $r_0$ to arrive at $\omega_e$ using standard result

$$\omega_e = (1/2\pi)\sqrt{(k_e/\mu)} = (1/2\pi)\sqrt{[e^2/(\mu r_0^3)]} \text{ s}^{-1} \qquad (4a)$$

we use the classical formula for spherical point-like particles with mass $m_x$, i.e.

$$m_x = (4\pi/3)\gamma_x r_x^3 \text{ g} \qquad (4b)$$

with $\gamma_x$, the density (g/cm$^3$) and $4\pi/3$, the spherical form factor. Macroscopic model (4b) for $m_x$ is reliable in classical physics. Questions emerge for micro-systems (a) *form factor and density*: is classical (4b) adequate for dihydrogen?; (b) *mass*: should total mass $2m_H$ or reduced mass $\frac{1}{2}m_H$ be used?; and (c) *size*: do results apply for $r_0$ in Coulomb energy $-e^2/r_0$ or for $2r_0$ in virial energy $-\frac{1}{2}e^2/r_0$?

The sum of electron and proton mass [26] leads to $m_H=1/(5,97538.10^{23})$ g. With $\gamma_H=1$, (4b) gives

$$r_H = [(3m_H/(4\pi\gamma_H)]^{1/3} \text{ cm} = 7,36516.10^{-9} \text{ cm} = 0,736516 \text{ Å} \qquad (4c)$$

as classical radius $r_H$, whereas Bohr gives $r_B=0,529177$ Å (without recoil). Since $r_{HH}=2r_H$, (4c) gives $r_0=r_{HH}=1,473032$ Å, typical for a virial rather than for a Coulomb energy (observed $r_0=r_{HH}=0,740144$ Å [25]). Even $\gamma_x=1$ is a fair[7] approximation for $H_2$ as a *dumbbell* ⊙⊙ with 2 spherical atoms at either side ($-\frac{1}{2}r_0$ and $+\frac{1}{2}r_0$) of the center of mass, which gives a *left-right anti-symmetric, achiral* configuration, as referred to in the title.

[Although a dumbbell is a standard model for a *rigid rotator* (diatomic bond), it has implications for vibrations. *Rotations of the complete dumbbell* (bond) are described with moment of inertia $I=mr^2$ and angular velocity $\omega$, giving $E_{rot}=\frac{1}{2}I\omega^2=\frac{1}{2}p^2/I$, with $p=I\omega=mvr$, using the center of mass of the dumbbell (one center system). As stated above, rotational J-states for $H_2$ are not considered here. However, *rotation-vibration coupling in a dumbbell* cannot do without degrees of freedom for its parts and without considering it as a 2-center system (non central fields).

---

[7] For systems with constant $m_x/\gamma_x$, all $r_x$ are (nearly) equal, as observed for isotopomers $H_2$, $D_2$ and $T_2$ [25, 27].

Rotation-vibration coupling in ⊙–⊙ can be visualized with a rope or twisted wire instead of a rigid bar to represent the inter-atomic field. This exposes symmetry details for atomic rotations at each center (contained in electronic term $E_0=E_n=E_{elec}$ and considered invariant to rotational symmetries). In fact, in a dumbbell model for atomic rotations perpendicular to field axis r, one atom must rotate *clockwise*, the other *anticlockwise* (or vice versa) to get at rope length variations by winding and unwinding[8]. If rotations were *clockwise*, the dumbbell is translated, as easily verified by looking at rotation as rolling[9] on a surface (*anticlockwise* rotation gives translation in the opposite direction). More generally, *atomic rotations* at each center in a dumbbell configuration are directed either *out-of-plane* (OP, i.e. perpendicular to field axis r) or *in plane* (IP, i.e. in a plane, containing r). All rotations are easily monitored by rolling.

- OP: with a fixed dumbbell center, disrotatory atomic rotation leads either to a rotation of the complete dumbbell or to oscillations of its parts along field axis r, as described above. Conrotatory motion (rolling without slipping) is prohibited with the dumbbell's center fixed. If free, conrotatory motion gives out-of-plane *translation*, without effect on internal symmetries.

- IP: with a free center, conrotatory motion displaces the dumbbell either to the right or to the left, which is also irrelevant for its internal dynamics. Only disrotatory motion (anti-symmetric spinning) will either make the gap between the spheres smaller (shorter bond) or larger (longer bond). The *linear velocity of the center of mass of rolling objects* (provoking an atomic displacement Δ) is equal to angular velocity times radius. This justifies the use of (3c) for $H_2$ and explains why *atomic vibrations* (3e) are linked to *anti-symmetric atomic rotations*. In a dumbbell view, rotation of one atom as a solid sphere is looked at from the other center. If so, this requires a H model, different than that in Bohr theory, to which we refer further below].

With this view on rotation-vibration coupling in a diatomic bond, we verify that

(i) with (4a) and (4c), the fundamental vibrational frequency[10] (in cm$^{-1}$) for $H_2$ becomes

$$\omega_e = 4410{,}1722 \text{ cm}^{-1} \tag{4d}$$

where 4402,93 cm$^{-1}$ [12] or 4401,213 cm$^{-1}$ [25] are observed;

(ii) with (4c), a virial energy for $H_2$ is

$$-V_0 = e^2/(2r_H) = a_0 = 78844{,}9125 \text{ cm}^{-1} \tag{4e}$$

(observed $a_0 = \frac{1}{2}k_e r_0^2 \approx 79000$ cm$^{-1}$ [28]). Since (4d)-(4e) have the same dimension, a *3$^d$ result* is that

(iii) a *natural quantum hypothesis* for bond $H_2$ emerges. The small ratio of elementary *step* $\omega_e \sim 4400$ cm$^{-1}$ (4d) and total gap $a_0 \sim 79000$ cm$^{-1}$ (4e) suggests that a number of successive integer *steps*, say v as in (1c), is needed to cover this gap. Since step and gap are both in cm$^{-1}$, the ratio is a number

---

[8] With one atom fixed in a rope model, the other rotates clockwise to fold; anticlockwise to unfold (or vice versa).

[9] Rolling, a useful motion on the macroscopic scale also appears at the microscopic, molecular or nano-scale, as proved with scanning tunneling microscopy (STM), to which we refer further below.

[10] The same formula for an electron ($m_e = m_H/1837{,}15267$ and radius $r_B$) gives $\omega_e = 219474{,}65 = 2*109737{,}31$ cm$^{-1}$, or twice the Rydberg $e^2/r_B$ [26]. This shows how the internal mechanics of H and $H_2$ are intimately connected.

$$q=\omega_e/a_0= 4410,1722/78844,9125=0,05593477 \tag{4f}$$

which can bring in quantization following *step* $\delta_v$, function of integer v (used to numerate the $H_2$ bands in the order they are observed [12]). The resulting *field quantum hypothesis for bonds* is

$$r/r_0-1=\Delta/r_0 =d_{HO}=\delta_v=qv \tag{4g}$$

Dimensionless (4g) must now be plugged into variable $d_{HO}$ and $d_{SK}$ for potentials $V_{HO}$ and $V_{SK}$. With (4f-g), $a_0\delta_v$ returns $a_0qv=v.4410,1722$ cm$^{-1}$ (see (1a-c). The inverse of coefficient q in (4g)

$$1/q=v_0=17,877967 \tag{4h}$$

is an internal maximum for v, i.e. for the observed bands for $H_2$. Since (4h) is higher than 14 (see Table 1), we will link this scale factor with scale factor f in (3f)-(3g) (see Section V).
With ionic Kratzer bond theory, the only input needed to solve the complete Hamiltonian for covalent bond $H_2$ and its oscillator (3p) is absolute mass of hydrogen atom $m_H$. Since $m_H \approx 1/N$ g (Avogadro $N=6,023.10^{23}$ [26]), $r_0$ for $H_2$ it assessable macroscopically and, as a result, $m_H$ provides with 3 fundamental parameters $\omega_e$, $r_0$ and $k_e$ for vibrator $H_2$, an unprecedented result. All vibrational characteristics for $H_2$ are now available in a classical, transparent way but we must find out how ad hoc quantum rule (4g) fits in old quantum theory.

*IV.3 Field quantum hypothesis for vibrations in bond $H_2$*

Neglecting recoil, angular velocity $v_e$ for a rotating electron $m_e$ is obtained from the ratio of *radial equilibrium condition* $m_e v_e^2/r=e^2/r^2$ with *quantum rule for angular momentum* $mv_e r=pr=n\hbar$, giving

$$v_e= m_e v_e^2 r/m_e v_e r=e^2/(n\hbar)= \alpha c/n \tag{5a}$$

similar to (3h). With Bohr radius $r_B$, quantized H-size differences[11] $\Delta_H$ become

$$r=e^2/m_e v_e^2=n^2\hbar^2/(m_e e^2)=n^2 r_B; \Delta_H=r-r_B=r_B(n^2-1) \tag{5b}$$

incompatible with linear quantum rule (4g). However, *if Bohr had quantized the Coulomb field as $e^2/n$ instead of angular momentum*, the same rotational energies $E_n=-R_H/n^2$ would have resulted, since

$$v_e=m_e v_e^2 r/m_e v_e r=(e^2/n)/\hbar \tag{5c}$$

is identical with (5a). Unlike (5b), quantum rule $e^2/n$ brings in a *linear* n-dependence

$$r=\hbar/(m_e v_e)=n\hbar^2/(m_e e^2)=nr_B \tag{5d}$$

instead of *quadratic* $n^2$ (5b). This gives a linear quantized difference on the field axis r

$$r-r_B=(n-1)r_B=\ell r_B \tag{5e}$$

whereby, instead of Bohr's quantum number n (with n=0 forbidden), Sommerfeld's secondary quantum number $\ell=n-1$ appears, where $\ell=0$ is allowed. With (5e), quantization for molecule $H_2$ proceeds through difference $\Delta$ between 2 separations on the field axis

$$\Delta_r=r-r_0=(n-1)r_0 \tag{5f}$$

---

[11] Difference $\Delta=r-r_0$ gives repetitions $r=r_0+\Delta=r_0+(r-r_0)=r_0+(r_0+\Delta)-r_0= r_0+r_0+(r-r_0)-r_0= r_0+r_0+(r_0+\Delta)-r_0-r_0\ldots$, easily avoided with (5f). For N repetitions $r/r_0=1+\Delta+N(+1-1)$, N *virtual pairs* (+1,–1) are created for a HO [29].

*linear*, instead of quadratic, in an *integer quantum number*. Its reduced dimensionless equivalent

$$\Delta_r/r_0 = r/r_0 - 1 = (n-1) = \ell \quad (5g)$$

provides with a Bohr-like validation of the above *field quantum hypothesis for vibrations in bonds* (4g) in a Hooke-Dunham $r/r_0$ theory. Since this differs from Kratzer's oscillator in $r_0/r$, a validation of (5g) depends on its implications for the $H_2$ spectrum.

## V. Quantization of symmetric linear and inverse field shifts in an achiral model

Multiplicative scaling in Kratzer's $r_0/r$ or Dunham's $r/r_0$ is additive. Inverse[12] and linear relations

$$\text{Kratzer: } r_0/r = r_0/(r_0 \pm \Delta) = 1/(1 \pm \Delta/r_0) = 1/(1 \pm \delta_r)$$

$$\text{Dunham: } r/r_0 = (r_0 \pm \Delta)/r_0 = (1 \pm \Delta/r_0) = (1 \pm \delta_r) \quad (6a)$$

where $\delta_r$ or $\delta_v$ is the numerical equivalent of a step, show quantization by (4g)-(5g). Rewriting total difference $\Delta$ between positions of 2 atoms on field axis r as

$$r - r_0 = +\Delta = +½\Delta - (-½\Delta) \text{ cm} \quad (6b)$$

reveals this is distributed in an *anti-symmetric way*, i.e. left and right to the center of mass, placed at the origin, but *equal in absolute magnitude* and based on the *arithmetic average*. In terms of symmetries, (6b) typifies an *achiral or too symmetrical $H_2$ bond* (as referred to in the title). Scheme (6a) shows why symmetric (6b) has different effects in Dunham and Kratzer models, as described below.

*V.1 The v-dependence in achiral mode: different analytical form of quantized Dunham and Kratzer oscillators*

(i) Symmetric distribution (6b), applied to Dunham's procedure for $r = r_0 \pm \Delta$ using (5g), gives

$$r/r_0 = 1 \pm \Delta/r_0 = (1 \pm \delta_r) = (1 \pm \delta_v) \quad (6c)$$

where left and right are avoided by virtue of (6b). Dunham's potential (2a) away from $r_0$ becomes

$$½k_e r_0^2 (r/r_0)^2 = a_0 (r/r_0)^2 = a_0 (1 \pm \delta_r)^2$$

With quantization rule (5g), reduced Dunham potential *differences* are

$$V'_{HO} - V'_0 = \Delta V'_{HO} = -½(1-qv)^2 + ½ = +qv - ½q^2v^2 \quad (6d)$$

Using $a_0$ (4e) and q (4f), the numerical result of achiral Dunham $H_2$ theory in cm$^{-1}$ is therefore

---

[12] Despite appearances, an additional classical constraint for differences between 2 so-called equal bonding partners $H_a$ and $H_b$ in dihydrogen $H_aH_b$ is available, if they are distinguished formally by mass $m_a$ and $m_b$ as well as by their positions on the field axis $r_a$ and $r_b$. As in a balance, reduced mass is based on classical

$$m_a r_a = m_b r_b (= C)$$

whereby C is a field dependent constant, with dimensions $(e/v)^2$. Dimensionless numerical equivalent relation $m_a/m_b = r_b/r_a$ suffices for recoil corrections. The underlying classical universal relations between $m_x$ and $r_x$ are

$$m_x = C/r_x \text{ or } r_x/C = 1/m_x.$$

noticing that $m_x r_x = e^2/v^2$ is consistent with (3b). If $r_{HH}$ required addition, reduced mass $\mu$ appears naturally, since

$$r_{HH} = (r_a + r_b) = C(1/m_a + 1/m_b) = C(m_a + m_b)/(m_a m_b) = C/\mu$$

Similarly, if total mass $m_{HH}$ required addition, reduced separation $\varrho = r_a r_b/(r_a + r_b)$ appears naturally too, since

$$m_{HH} = m_a + m_b = C(1/r_a + 1/r_b) = C(r_a + r_b)/(r_a r_b) = C/\varrho$$

This explains the difficulties above with (4c), the classical result for $r_{HH}$, since $\varrho = ½r_{HH}$ for $H_2$.
If the sum-based reduced separation is $\varrho_+$, a difference-based reduced separation $\varrho_-$ obeys

$$1/\varrho_- = 1/r_a - 1/r_b = (r_b - r_a)/r_a r_b$$

to which we return further below, see (6f).

$$\Delta V_{HO} = \Delta E_v = 4410{,}17v - 123{,}34v^2 \text{ cm}^{-1} \qquad (6e)$$

close to a 2$^{nd}$ order fit[1] (Section II) but with large errors of 111 cm$^{-1}$. The improvement over Schrödinger's HO (1a) may be considerable, *spectroscopic accuracy* is far away. Morse's quadratic in (v+½) is only moderately successful too [7,14]. A parameter for qv cannot improve a fit.

(ii) To apply field quantization for a Kratzer potential, there is a problem with anti-symmetric or left-right symmetric distribution (6b). Inverse[12] $r_0/r = r_0/(1\pm\Delta)$ in (6a) does not account for the positions of 2 atoms $H_A$ and $H_B$ with respect to the center, i.e. *achiral* distribution $\pm½\Delta$ in (6b). To understand this, we return to the different equivalent rearrangements of four inter-atomic Coulomb terms, see Section IV between (3a) and (3b), where generalized Coulomb term[12]

$$(e^2/r_0)(r_0/r_A - r_0/r_B)$$

appears, a composite Coulomb term of Kratzer type. Field quantization with Kratzer's variable $r_0/r$ therefore uses refined radial variables, obeying respectively

$$r_A = r_0 - ½\Delta \text{ and } r_A = r_0 + ½\Delta$$

due to positional symmetry (achiral system). In exactly the same way as the generalized Coulomb term above, the Kratzer-Coulomb variable now becomes

$$r_0/r_A - r_0/r_B = r_0(r_B - r_A)/r_A r_B = 1/(1-½\delta_r) - 1/(1+½\delta_r) = \delta_r/(1-¼\delta_r^2) \qquad (6f)$$

Using (5g), the quantized v-dependence for this Kratzer variable is

$$1/(1-½qv) - 1/(1+½qv) = qv/(1-¼q^2v^2) \qquad (6g)$$

instead of linear qv in Dunham's (6d). The reduced Kratzer oscillator difference is

$$\Delta V'_{SK} = -½[1 - qv/(1-¼q^2v^2)]^2 + ½ = +qv/(1-¼q^2v^2) - ½q^2v^2/(1-¼q^2v^2)^2 \qquad (6h)$$

to be compared with Dunham's (6d). In cm$^{-1}$, the numerical Kratzer result is

$$\Delta V_{SK} = (+4410{,}17v - 123{,}34v^2 - 3{,}49v^3)/(1 - 0{,}00078v^2)^2 \text{ cm}^{-1} \qquad (6i)$$

the performances of which are discussed in the next section. Relation (6i) entails naturally higher order terms in v, suggested by (1b)-(1c), to accommodate for anharmonicity. Unlike (6c), a parameter for qv in (6g) can affect the goodness of fits. As for (6e), also (6i) is an analytical first principles' formula of closed form, based solely on $m_H$ as input for the complete $H_2$ spectrum. Maximum $v_0$ (4h) derives from *variable* qv, rewritten as $v/v_0$. For *Dunham oscillators* $(1-x)^2 = (1-v/v_0)^2$, $x = v/v_0 = 0$ returns the well depth; $x=1$ or $v=v_0$ gives zero. For *Kratzer oscillators* $(1-y)^2$, $y = (v/v_0)/[1-¼(v/v_0)^2] = qv/[1-¼(qv)^2]$, $y=1$ implies that $qv = 1 - ¼(qv)^2$ or $v^2 + 4v/q - 4/q^2 = 0$. Solving for $v_0$ gives

$$v_{0(KR)} = v_{0(DU)}/[½(1+\sqrt{2})] = 17{,}877967/1{,}207107 = 14{,}810593 \qquad (6j)$$

in line with 14 observed levels (Table 1). The fact that band 15 is missing is another unprecedented result of Kratzer theory. This is confirmed[7] by the greater number of bands for $D_2$ and $T_2$ [25,27].

*V.2 Results with quantized Dunham and Kratzer oscillators*

Since optimization is used widely in QM, a parameter for qv is allowed. A multiplicative or external parameter $p_e$ cannot improve the goodness of a fit, since size does not affect classical Euclidean symmetries (ratio's, proportions). However, internal parameters $p_i$ affect (dynamic) symmetries. In parameterized HO $[p_e(x_1-p_i x_2)]^2$, the position of the extreme is not affected by $p_e$ but it is by $p_i$. Whereas external $p_e$ cannot affect the goodness of a fit for a vibrator, internal $p_i$ can. Typical *external scaling parameters* for bonds are Dunham's $a_0$, fundamental frequency $\omega_e$, bond energy $D_e$, all in cm$^{-1}$, if energy E(r) is in cm$^{-1}$. Non-dimensionalization with *external multiplicative scaling* parameters[13] only generates new variables, commensurate with these scale factors. *Internal parameters* can determine the goodness of a fit as they refer to *internal or dynamical symmetries*.

To normalize results, we compare variable qv or Dunham's $\delta_{HO}$ (for which parameterization is ineffective), with parameterized Kratzer's $\delta_{SK}/p$ (p being an internal parameter $p_i$) using

(i) $\quad \delta_{HO}=qv$ \hfill (7a)

(ii) $\quad \delta_{SK}/p=(1/p)[1/(1-½pqv)-1/(1+½pqv)]= qv/(1-¼p^2q^2v^2)$ \hfill (7b)

This secures leading term qv is identical for all 14 vibrational levels v in *either method*. The main difference between the 2 resides in normalizing factors: 1 for Dunham's but $1/(1-¼p^2q^2v^2)$ for Kratzer's potential, although critical points can emerge because of $1/(1-½pqv)$. Normalizing Kratzer's potential as in (7b) brings in *harmonic mean* $[(1-½pqv)(1+½pqv)]=(1-¼p^2q^2v^2)$, a more natural feature to discuss a *harmonic* oscillator.

The accuracy of the 2$^{nd}$ order fit with Kratzer's variable (7b) is maximum for $p=p_i=0,83795$. The 2$^{nd}$ order fits for plots of levels versus $\delta_{HO}$ (7a) and $\delta_{SK}/0,83795$ (7b) in Fig. 2 are respectively

$\quad E_{\delta(HO)}=-40971,3574\delta_{HO}^2+78614,1312\delta_{HO}-161,1126$ cm$^{-1}$ \hfill (7c)

with goodness of fit $R^2=0,9998627$ and

$\quad E_{\delta(SK)}=-40754,1814\delta_{SK}^2+76766,2419\delta_{SK}-3,56576$ cm$^{-1}$ \hfill (7d)

with $R^2=0,9999999$.

Although coefficients in (7c) and (7d) are comparable with values as theoretically expected, their difference clearly shows in the errors given in Table 2, see also Fig. 3. Kratzer's errors of 3 cm$^{-1}$ or 0,021 %[14] for (7d) almost vanish when compared with Dunham's: they are 30 times smaller than for (7c), i.e. 111 cm$^{-1}$ or 0,54 %. Errors for (7d) are 530 times smaller than Schrödinger's HO recipe (1a) with errors of 1840 cm$^{-1}$ (see Section II). Also, errors of 3,4 cm$^{-1}$ for *simple Kratzer bond theory* are equal to those of a *complicated, early ab initio QM method* [10], i.e. 3,2 cm$^{-1}$, cited by Dabrowski [12]. Kratzer's simple 2$^{nd}$ order parabola is even more accurate than Dunham's a 4$^{th}$ order fit in v, with its errors of 7 cm$^{-1}$ (Table 2). A 4$^{th}$ order Dunham oscillator has the 3 terms $d_D^2$, $d_D^3$ and $d_D^4$ in (2c). The accuracy of a 4$^{th}$ order fit with the same Kratzer variable is not significantly better (not shown). This is rather surprising, since a less symmetrical (chiral) structure should obey a Hund-type double well curve (a quartic, 4$^{th}$ order in v)[31,32]. If H$_2$ were chiral, *left-right asymmetry* instead of *left-right symmetry* (6b) must show. This left-right problem for H$_2$ was discussed in Section IV.2 with the distinction between dis- and conrotatory atomic motion. As soon as such effects are exposed in a spectrum, *atomic anti-symmetry* appears. If one atom in the dumbbell belonged to a *left-handed* frame, the other must belong to a *right-handed* one or, *if one partner in a diatomic bond were an atom, the other must be a charge-inverted anti-atom* [22]. To account for left-right asymmetry in a generic way, we can use either a geometry dependent parity operator **P** or a geometry independent charge-inversion operator **C**, an important dilemma referred to in Section VI.

---

[13] QM parameterization is typically multiplicative or external. This was criticized in the EPR-paper [30] on the completeness of QM: only additive scaling can affect symmetry-effects associated with variables.

[14] Including atom energies (1 Hartree) and covalent $D_e$ (sum 246500 cm$^{-1}$), % errors are artificially reduced to 0,0015. For 14 bands between ~90000 and ~55000 cm$^{-1}$ [12], % errors in this work would be equal to only 0,011.

Here, we discuss a last technical but equally important problem: how to assess analytically the H$_2$ *covalent bond energy* D$_e$ from an *ionic* Kratzer potential or from the *ionic bond energy* D$_{ion}$.

*V.3 Covalent H$_2$ bond energy D$_e$ from an ionic Kratzer potential or D$_{ion}$*

Oscillator D(1-x)$^2$ and oscillator difference D(1-(1-x)$^2$)=D(2x-x$^2$) transform in D((1-x')-1)$^2$=Dx'$^2$ and D[(1-x')$^2$-2(1-x')] with complementary[15] variable x'=1-x. for the latter, a plot versus x' gives well-depth D as an intercept, since the linear term *has vanished* with the complementary variable[11]. Although Coulomb's –e$^2$/r vanishes exactly *by this complementary variable, one cannot conclude that the system is not of Coulomb-type or not ionic.* For the better performing Kratzer potential (7d) for H$_2$, its first derivative d/dδ$_{SK}$ (or d/dδ after dropping the suffix) gives extreme δ$_{max}$=0,9418204. The maximum well depth, i.e. the *covalent bond energy* D$_e$ of H$_2$, is therefore

D$_e$= 36146,442 cm$^{-1}$

Complementary unit +1=+x+(1-x) is now +δ$_{max}$=+δ +(δ$_{max}$- δ). Scaling with δ$_{max}$=0,9418204 gives a complementary unit description in Kratzer variable δ, applicable for H$_2$, i.e. +1=1,0617773521δ +1 -1,0617773521δ. External parameter p$_e$=1/ δ$_{max}$ makes first order Coulomb term vanish exactly. Fig. 4 shows level energies plotted versus x=p$_e$δ and x'=(1-x)=(1-p$_e$δ). The 2$^{nd}$ order fits are respectively

E$_x$ = -36150,0077x$^2$ + 72300,0154x - 3,5658 cm$^{-1}$   (8a)

wherein 72300,0154≡2.36150,0077 as required and

E$_{x'}$= -36150,0077x'$^2$ + 0,0000x' + 36146,4419 cm$^{-1}$   (8b)

giving errors as reported in Table 2. *Ionic Kratzer potential* (8b) gives intercept D$_e$=36146,44 cm$^{-1}$, within 0,078 % of observed D$_e$=36118,3 cm$^{-1}$ (without zero point energy [28]), which proves that *ionic Coulomb attraction* -e$^2$/r is at work in *covalent* H$_2$, treated as an *achiral* system (6b). While this theory has the same first principle's status as Bohr H theory, its results are much better than with Schrödinger's (1a) and with Heitler-London theory[16] [33], published immediately thereafter.

---

[15] Complementarity +1=+x+(1-x) is valid, however x is defined. Since any x will do, this equation is useless, if not trivial, unless constraints can be imposed (see text).
[16] Heitler and London obtained less accurate r$_0$=0,80 Å, ω$_e$=4800 cm$^{-1}$ and D$_e$=3,14 eV or 25300 cm$^{-1}$ [33].

*V.4 Formal connection with Bohr H theory*

When compared with (1a), an advantage of (8b) is that average 36148=½(36150+36146) gives

$$E_{x'} \approx 36148(1-x'^2) = D_e(1-x'^2) \text{ cm}^{-1} \qquad (8d)$$

as simplified *ionic* Kratzer band equation, with asymptote *covalent* $D_e$, for *a complete molecular band spectrum* ($H_2$) based on v-quantization. This equation is formally similar to Bohr's formula

$$E_n = R_H(1-1/n^2) \text{ cm}^{-1} \qquad (8e)$$

for *a complete line spectrum* (H Lyman series), with Rydberg $R_H$, based on n-quantization [34]. A simple *ionic* Kratzer bond theory makes *covalent* bond $H_2$ prototypical for molecular spectroscopy, just like simple Bohr theory made atom H prototypical for atom spectroscopy (see Introduction).

## VI. Discussion

(i) *Ionic bond theory* rationalizes *covalent bond* $H_2$. Interactions $H^+H^-$ and $H^-H^+$ are typified with *ionic Coulomb attraction* $V=-e^2/r_{AB}$. Particle transfers have no effect on total mass since $2m_H \equiv (m_H-m_e)+(m_H+m_e)$ and only a negligible effect for *reduced mass*: $\mu_{ion}=½(m_H-m_e)(m_H+m_e)/m_H=½m_H[1-(m_e/m_H)^2]$ differs from $½m_H$ by only $3.10^{-5}$ %. *Resonance* between $[H^+H^-;H^-H^+]$ at $r=r_0$ avoids a permanent dipole moment for $H_2$ [22]. However, other problems emerge.

(ii) Dis- or conrotatory motion of 2 neutral atoms with the same charge distribution is a first problem. The left-right distinction, connected with disrotatory motion, leads to a bond between an atom and an antiatom, see Section V.2, whereby positions (coordinates) are not affected. Although anti-symmetry is usually approached by spin symmetries, it is well known from H line spectra that spin effects $\sim m_e\alpha^4c^2$ are $\alpha^2$ (or $137,04^2$) times smaller than energies $\sim m_e\alpha^2c^2/n^2$ [16,34]. Spin can never generate a switch from *mutually exclusive repulsive* $+e^2/r_0$ to *attractive* $-e^2/r_0$, conform the $H_2$ band spectrum nor can a permutation of positions, unless charges are interchanged too.

(iii) The next problem, related with (i) and (ii), is the value as well as the sign of numerical field form factor $A_r$ in (3d). Why must *attractive ionic* $V=-e^2/r_{AB}$ be used, while, in (3a), *mutually exclusive repulsive* $+e^2/r_{AB}$ appears instead? Also here, a simple mathematical solution exists but this creates problems for physicists[17], some of which are still unsolved today [22,28,32,34].

As argued in Section V.2, this solution allows *Coulomb attraction* $-e^2/r_{AB}$ between neutral atom H and its *charge-inverted* partner H̲ [22], by virtue of forbidden charge operator **C**. We showed in Section IV.1 that **C** seems competitive with parity operator **P**. By definition, atom and anti-atom always react in an anti-symmetric way towards a field of Coulomb or electromagnetic nature, whatever the geometry of the structure to which they belong. If geometry dependent form factor $A_r$ remained constant, which is like saying that, while the size can change, geometry cannot, a switch of sign for constant $A_r$ can only find its origin in operator **C**.

---

[17] One of these problems is the phenomenological extension of Dirac annihilation between a pair of *charged (charge-conjugated) elementary particles* towards a pair of neutral atomic species like H and H̲. This extension is far from evident with the Coulomb forces in pairs *of 2 charge-conjugated non-composite particles* and *of 2 neutral composite particles*.

(iv) C is given away by the $H_2$ spectrum but is conventionally forbidden in the natural world. It brings in H<u>H</u>; <u>HH</u> interactions between 2 neutral atomic species, which create major problems for physics. If allowed, they would at least also solve some longstanding problems and they certainly would make the theory of the chemical bond more transparent than QM [22]. In practice, the effect of **C** for a bond is limited to the 4 inter-atomic Coulomb terms in (3a), which applies for both HH and <u>HH</u> by **C**-symmetry [22]. However, E(H<u>H</u>) leads directly to

$$E(H\underline{H})=\tfrac{1}{2}m_a v_a^2+\tfrac{1}{2}m_b v_b^2+\tfrac{1}{2}m_A V_A^2+\tfrac{1}{2}m_B V_B^2 - e^2/r_{aA} - e^2/r_{bB} -(-e^2/r_{bA} - e^2/r_{aB} + e^2/r_{ab} + e^2/r_{AB})$$
$$=\tfrac{1}{2}m_a v_a^2+\tfrac{1}{2}m_b v_b^2+\tfrac{1}{2}m_A V_A^2+\tfrac{1}{2}m_B V_B^2 - e^2/r_{aA} - e^2/r_{bB} + e^2/r_{bA} + e^2/r_{aB} - e^2/r_{ab} - e^2/r_{AB}$$

with internucleon attraction $-e^2/r_{AB}$ instead of repulsion $+e^2/r_{AB}$ in (3a) [22]. This transformation being independent of the system's geometry, **C** is generic: forbidding it a priori on purely theoretical grounds is, to say the least, debatable[17]. Moreover, looking at the discussion around (3e) in Section IV.1, H<u>H</u> interactions not only lead to *attractive* $-e^2/r_{AB}$ as required with classical physics, but also to *real momentum*, whereas the HH interactions of QM lead to *repulsive* $+e^2/r_{AB}$ and *imaginary momentum*. Which scheme is the better must be decided with further work on H and $H_2$. Many consequences of **C** having been discussed in [22,28,32,34], we proceed with other points.

(v) Kratzer Coulomb energy $-e^2/r_0$ is important for universal behavior and the UF [2,23]. Scaling by *ionic bond energy* $D_{ion}$, rather than *covalent* $D_e$ [7,23] unifies the spectroscopic constants of *ionic and covalent bonds* between *all monovalent atoms in the Table* [2,23,35]. Difficulties, generated by scaling without $D_e$ [36] illustrates similar shortcomings of Dunham theory, like those exposed here.

(vi) The fact that *ionic bond energy* $D_{ion}$ can be a better scaling aid [2,23,35] than *covalent* $D_e$ has now been rationalized with an analytical relation between $D_{ion}$ and $D_e$ (see Sections V.4-5).

(vii) Universal behavior is usually connected with the smooth G(F)-plot of functions F for $\alpha_e$ and G for $\omega_e x_e$, whereby F and G relate to Dunham coefficients $a_1$ and $a_2$ in (2c) and to variable $r/r_0$. With a Kratzer parabola in $r_0/r$, higher order terms are superfluous; higher order terms in v are only generated by the connection between v and $r_0/r$ as in (6i). With (4d)-(4e), quadratic Kratzer term $\tfrac{1}{2}(e^2/r_0)(\omega_e/a_0)^2 = 0{,}5*4410{,}17^2/78844{,}91 = 123{,}34$ cm$^{-1}$ is in agreement with observed $H_2$ levels. This $2^{nd}$ order Kratzer term is close to $H_2$ anharmonicity $\omega_e x_e$ of 123,07 cm$^{-1}$ [6,12,25] in Dunham theory, where it is related to the $4^{th}$ order term with coefficient $a_2$ [2,7].

(viii) While Morse and Dunham theories are used more widely than Kratzer's[18] [2], the interest in Kratzer's function [7,8,13,37] is justified as it connects rotation (rolling[9] [38]) and vibration.

(ix) Double photoionization of $H_2$ [39] confirms the importance of non-Heitler-London *ionic states* for the $H_2$ ground state, which is exactly the result of *ionic* Kratzer bond theory [40].

---

[18] Applications of Kratzer's potential to other fields, e.g. nuclear physics, are not discussed here.

(x) For isotopomers HD, $D_2$… results must be as accurate as for $H_2$, since, even in simple approximation $m_D=2m_H$, similar $r_0$ values are obtained for $D_2$. This suffices to extend the ionic theory to covalent isotopomers[6] [25,27], without having to give details here.

(xi) With ionic Kratzer bond theory for covalent $H_2$ (6j), it is also readily understood why only 14 bands are observed for the $H_2$ spectrum. In essence, this maximum value of 14,81 (6j) derives from solely from mass $m_H$ too, as argued above. The complete $H_2$ bond theory can therefore be rewritten in terms of a numerical variable $v/v_0$, still having an ionic mechanism at its basis.

(xii) We do not expand on possible implications for metrology and the constants, instigated by new first principles relations like $\omega_e/e=(\tfrac{1}{2}m_H r_0^3)^{-1/2}$, $e/\omega_e=\sqrt{(2\pi/3)}r_0^3$…, generated in this work. Our work is formally consistent with classical moment[12] $m_x r_x = C = (A_r e^2/v^2)$, real by definition. The ratio of Schrödinger's imaginary momentum ip with real classical moment, i.e. $im\omega r/(mr)=i\omega$ leaves us with imaginary angular velocity and imaginary frequency. Any theory for real systems, like Schrödinger's, based solely on imaginary momentum, is in contradiction with observation: real systems only lead to real spectra, with real frequencies.

(xiii) Despite the good performances of an *ionic* Kratzer-Coulomb oscillator for *covalent* $H_2$, its relatively small errors[11] are not of spectroscopic accuracy[19] but are comparable with those of earlier QM calculations [10], respectively 3,4 (Table 2) and 3,2 cm$^{-1}$ [10]. In both cases, errors are much larger than with more elaborate relativistic QM, bearing on nonadiabatic corrections [3]. However, Kratzer theory needs only one *parameter* for optimization, whereas QM [3,4,10] needs many. Kratzer's old-quantum theory gives acceptable results without a wave equation, whereas QM methods [3,10] need hundreds of terms in the wave function of the simplest bond of all, $H_2$. This illustrates a few conceptual and computational advantages of Kratzer-oscillator bond theory.

## VII. Conclusion

A simple, reasonably accurate bond theory exists, in line with the existence of a UF. This *ionic* Kratzer bond theory treats *achiral covalent* bond $H_2$ different from conventional Dunham theory. It gives an analytic connection between *ionic and covalent bond energies*, whereby only hydrogen mass $m_H$ is needed as input. These unprecedented results justify a search for *a more accurate, less symmetrical or chiral, ionic Kratzer bond theory*, which we present later [21].

Schrödinger's choice to interpret the Hamiltonian as an energy operator leads to a complex theory for the chemical bond. This complexity is avoided in old quantum theory without loss of accuracy. For the theory of the chemical bond and as far as accuracy for $H_2$ is concerned, we safely conclude that the Bohr-Sommerfeld-Compton-de Broglie recipe to replace momentum $p_x=mv_x$ by $\hbar/r_x$ remains as valid as Schrödinger's to replace it by operator $(\hbar/i)\delta/\delta x$. These conclusions are validated by conceptual, theoretical and computational advantages of ionic Kratzer bond theory over QM theories like that of Heitler and London, as argued earlier in a different context [22,23]. Since *intra-atomic anti-symmetry* is probed by the band spectrum, atom-antiatom bonding may also dispose of the longstanding mystery of the matter-antimatter asymmetry in the Universe [22]. Results of spectroscopic accuracy for $H_2$ will be presented shortly [21].

---

[19] $H_2$ bands are accurate to 0,1 cm$^{-1}$ [12,41] or 3000 MHz, which is less accurate than for H lines [34].

Table 1. Observed vibrational levels for $H_2$ [12] (in cm$^{-1}$)

| v | quanta | levels |
|---|--------|--------|
| 0 | 4401,21 | 0,00 |
| 1 | 4161,14 | 4161,14 |
| 2 | 3925,79 | 8086,93 |
| 3 | 3695,43 | 11782,36 |
| 4 | 3467,95 | 15250,31 |
| 5 | 3241,61 | 18491,92 |
| 6 | 3013,86 | 21505,78 |
| 7 | 2782,13 | 24287,91 |
| 8 | 2543,25 | 26831,16 |
| 9 | 2292,93 | 29124,09 |
| 10 | 2026,38 | 31150,47 |
| 11 | 1736,66 | 32887,13 |
| 12 | 1415,07 | 34302,20 |
| 13 | 1049,16 | 35351,36 |
| 14 | 622,02 | 35973,38 |

Table 2 Errors for $H_2$ levels with Dunham (2nd and 4th order) and Kratzer (2nd order) functions (in cm$^{-1}$)

| v | levels | Dunham | | Kratzer |
|---|--------|--------|--------|---------|
|   |        | 2nd order | 4th order | 2nd order |
| 0 | 0,00 | 161,11 | 8,08 | 3,57 |
| 1 | 4161,14 | 53,18 | -8,34 | -3,91 |
| 2 | 8086,93 | -33,74 | -8,43 | -3,94 |
| 3 | 11782,36 | -94,63 | -1,67 | -0,77 |
| 4 | 15250,31 | -126,64 | 5,20 | 2,45 |
| 5 | 18491,92 | -128,61 | 8,56 | 4,15 |
| 6 | 21505,78 | -101,96 | 7,09 | 3,59 |
| 7 | 24287,91 | -50,66 | 1,75 | 1,13 |
| 8 | 26831,16 | 18,13 | -4,81 | -2,02 |
| 9 | 29124,09 | 92,98 | -9,39 | -4,48 |
| 10 | 31150,47 | 157,65 | -8,71 | -4,51 |
| 11 | 32887,13 | 188,98 | -1,59 | -1,37 |
| 12 | 34302,20 | 155,08 | 9,30 | 4,04 |
| 13 | 35351,36 | 11,66 | 13,69 | 6,80 |
| 14 | 35973,38 | -302,54 | -10,72 | -4,73 |
| **absolute error** | | **111,84** | **7,15** | **3,43** |
| **% error** | | **0,536** | **0,044** | **0,021** |

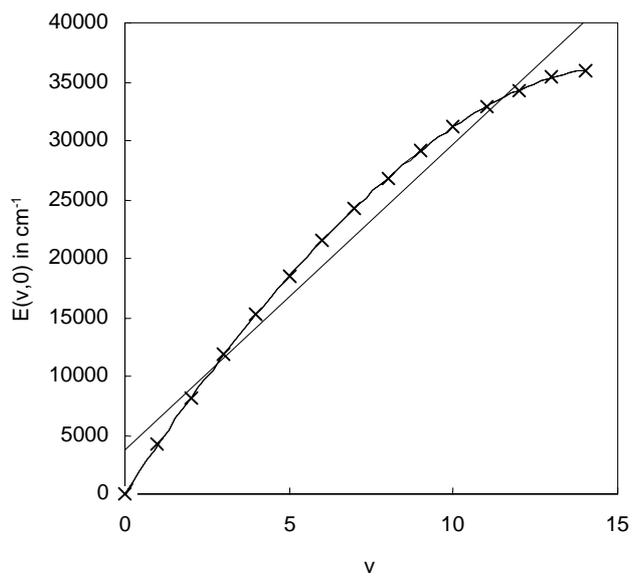

Fig. 1 Plot of 14 vibrational levels E(v,0) versus v [12].
Linear fit (full line); 2nd, 4th and 6th order fits coalesce to a single broad curve (dashes).

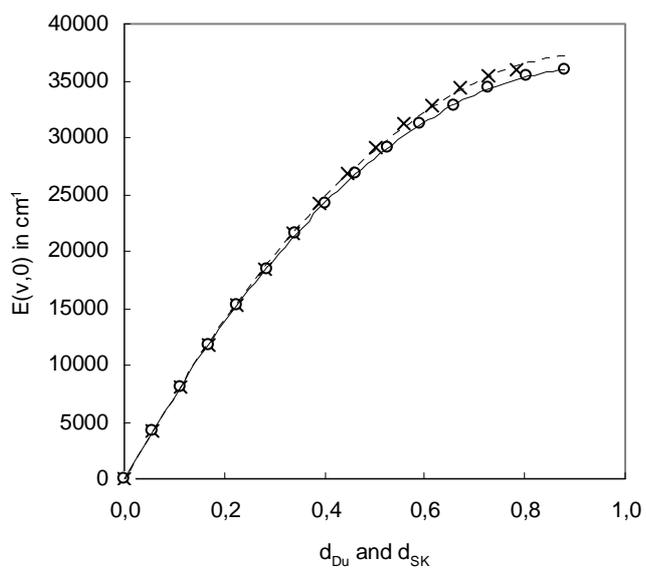

Fig. 2 Plot of E(v,0) versus $d_{DU}$ (dashes) and $d_{SK}$ (full) for 2nd order fits

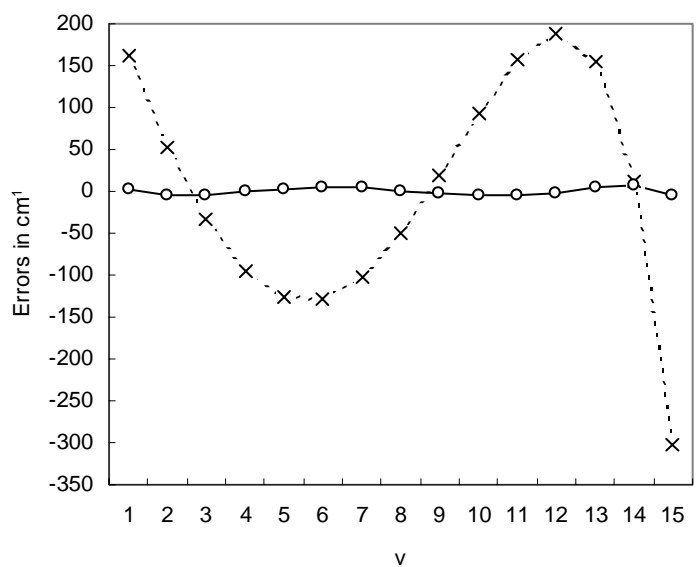

Fig. 3 Errors with 2nd order fits for Dunham (x) and Kratzer (o) oscillators.

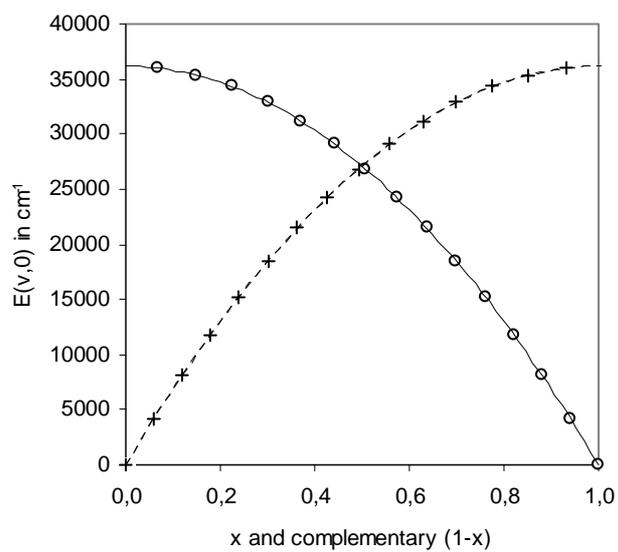

Fig. 4 Energy levels with Kratzer parabola (8a) versus x (+, dashes) and
(8b) versus complementary 1-x (o, full line), giving $D_e$ as intercept (see text)